\documentclass{article}
\usepackage{cite}
\usepackage{amsmath,amssymb,amsfonts}
\usepackage{algorithmic}
\usepackage{graphicx}
\usepackage{textcomp}
\usepackage{authblk}
\usepackage{xspace}

\usepackage{xcolor}
\def\BibTeX{{\rm B\kern-.05em{\sc i\kern-.025em b}\kern-.08em
    T\kern-.1667em\lower.7ex\hbox{E}\kern-.125emX}}
\DeclareUnicodeCharacter{2212}{-}

\begin{document}

\title{A Note on Look Ahead in Real Life and Computing}
\author{*Burle Sharma, **Rakesh Mohanty}
\affil{Department of Computer Science and Engineering\\Veer Surendra Sai University of Technology\\Burla, Odisha, India\\*burlesharma,**rakesh.iitmphd@gmail.com }

\author{Sucheta Panda}
\affil{Department of Computer Application\\Veer Surendra Sai University of Technology\\Burla, Odisha, India\\pandasucheta@gmail.com }

\maketitle

\begin{abstract}
    Past, Present and Future are considered to be three temporal and logical concepts which are well defined by human beings for their existence and growth.   We, as human beings, have the privilege of using our intelligence to mentally execute an activity before physical occurrence of the same in the real world.  Knowledge of the past, aplomb of present and visualisation for the future correspond to three concepts such as look-back, look-at and look-ahead respectively in real life as well as in diversified domains of computing.  Look-Ahead(LA) deals with the future prediction of information and processing of input to produce the output in advance.  In this article, our main objective is to learn,  understand and explore the concept of LA and design novel models as solution for real world problems.  We present three well known algorithmic frameworks used in practice based on availability of input information such as offline, online and semi-online.  We introduce interesting real life applications and well known computing problems where LA plays a significant role for making a process, system or algorithm efficient. We define new types of LA and propose a taxonomy for LA  based on literature review for designing novel LA models in future. Using the concept of LA, We identify and present many interesting and non-trivial research challenges as future potential research directions.  Intuitively, we observe that LA can be used as a powerful tool and framework for future researchers in design of efficient computational models and algorithms for solving non-trivial and challenging optimization problems.   
\end{abstract}
\emph{Keywords -
Look-Ahead(LA), optimization, online, offline, semi-online, competitive analysis}  

\tableofcontents
\newpage

\section{Introduction}

Look-Ahead(LA) is one of the ubiquitous concepts which affects significantly the outcomes of various activities in our day to day life.  LA deals with future prediction of input information and output. We make better decisions in our life by visualizing and planning for the future.  Setting of any goal and attaining an objective to accomplish a task involves planning with looking ahead of time.  

The concept of LA can be incorporated in our diet plan in day to day life. After getting up early in the morning, we plan for our breakfast as a healthy diet for the day then we plan our lunch ahead of time.  Similarly, we plan our evening snacks and dinner well before time.  We as human beings make our diet plan for the day ahead of time before occurrence of the event such as breakfast, lunch and dinner for healthy living.   

Every individual's life is based on three phases of time-lines such as past, present and future.  The past can be mapped to the concept of look-back, present to the concept of look-at and future to the concept of look-ahead.  The activities already performed beforehand becomes a history of the past as look-back.  In many emergency situations we instantly make decisions at present by using the concept of look-at.  To increase our efficiency and for productive outcome, we think, plan and act ahead of time by using the concept of LA.  The decisions made in the past cannot be revocable whereas the decisions to be made in present and future can be adaptable.  

It is interesting and challenging to raise the following questions for further investigation.
\begin{itemize}
    \item What is the role of LA in design of novel Computational Models for challenging real life problems?
    \item What are the roles of Past, Present and Future inputs in Algorithm Design?
    \item Why should we focus on present and future inputs for designing efficient Computing Systems?
\end{itemize}
  
LA concept has also been extensively used in computational and optimization problems in the literature 
\cite{albers1998competitive}.

\subsection{Optimization Problems}

Optimisation\cite{albers1998competitive,dunke2014online, kukunuri2019lookahead} is the process of obtaining best solution from a set of feasible solutions.  The main aim of optimisation is to minimize the cost or maximize the profit.  For instance, before we buy vegetables, we prepare a list of items to be purchased and make a plan to visit multiple vegetable shops in the whole market to learn about the price of vegetables before hand to minimize the cost. In the manufacturing sector, people strive to manufacture high-quality goods with cheaper cost. An airline company needs to optimize its flight schedules to minimize fuel costs while meeting passenger demand and crew scheduling constraints.  A logistics company must optimize its transportation routes to minimize fuel consumption and transportation costs while delivering goods to various locations.  A mutual fund manager seeks to minimize transaction costs, such as trading fees and taxes, while rebalancing a fund's portfolio.  A homeowner aims to minimize energy costs by optimizing the use of heating and cooling systems, insulation, and renewable energy sources. 
 A package delivery service aims to minimize delivery times and fuel costs by optimizing delivery routes and vehicle assignments.  A marketing department looks to minimize advertising costs while achieving the desired level of brand exposure and customer engagement.

A shopkeeper plans and makes more money by selling fruits, vegetables, and groceries for maximizing profit.  A student strives to achieve highest possible grades  in an examination by following efficient strategies for maximizing the score. A manufacturing company must decide how many units of a product to produce to maximize its profit while considering factors like production costs, demand, and selling price. An investor aims to maximize their portfolio's return while considering risk tolerance, asset allocation, and investment choices. A power plant operator wants to maximize profit by optimizing the generation of electricity while minimizing fuel consumption and emissions. A ride-sharing company seeks to maximize driver profits by matching drivers with passengers efficiently, considering factors like demand, distance, and driver availability.A chemical plant manager wants to maximize the production of a high-value product while minimizing waste and energy consumption. An e-commerce company aims to maximize profit by optimizing online advertising spending, targeting the right audience, and adjusting pricing strategies.

Optimization in practice is challenging, because we lack complete information about the input.  Finding efficient solution to an optimization problem under uncertainty is most non-trivial and challenging. To solve the optimisation problem we design efficient algorithms.

\subsection{Offline and Online Framework}
Based on availability of input information the algorithm can be classified as
 online and offline \cite{bachrach2002competitive, borodin2005online, reingold1996off, reingold1990optimum}.  When using an offline algorithm, the entire set of input is available at the outset whereas in online algorithm only the current and past inputs are available at any instant of time without knowledge of the future inputs.  
Selection sort\cite{hester1985self, sleator1985self} is an example of offline frame work and insertion sort is an example of online frame work. 

The selection sort algorithm organizes an array by iteratively identifying the smallest element (in ascending order) from the unsorted portion and relocating it to the start of the array. This method necessitates complete access to the entire input data set. Conversely, insertion sort evaluates a single input element during each iteration, generating a partial solution without contemplating future elements. Consequently, insertion sort qualifies as an online algorithm. In the context of online algorithms, LA refers to the ability to anticipate future input data. An online algorithm possesses LA of zero, meaning it operates solely on current data without forecasting, while an offline algorithm has a finite LA restricted by the input size.

We present the offline vs online framework in Table 1 as follows

\begin{table}[h]
    \centering
    \begin{tabular}{|c|c|}
    \hline
         offline & online \\
        \hline
        Complete input information & Incomplete input information \\
        \hline
        Whole input sequence available & inputs on the fly \\
        \hline
        visibility of complete future & visibility of instant future \\
        \hline
        performance better than online & performance worse than offline \\
        \hline
        sometimes far from reality & more realistic\\
        \hline
    \end{tabular}\\
     \caption{Offline vs offline comparison}
    \label{tab:my_label}
\end{table}
\subsection{Offline, Online and Semi-Online}
An online algorithm is a decision-making process where we have to make choices one by one, without knowing the future. It is like playing a game where we cannot see the next moves. Instead, we make decisions as we go, often with limited information.

Imagine we are baking cookies, but we do not have the recipe in advance, and we cannot see all the ingredients in the kitchen. We  have to make decisions about which ingredients to use and how much of each one as we discover them.

In the online algorithm context:

\begin{itemize}
    \item We receive inputs or data one at a time, and  make  decision immediately.
    \item We can't change your previous decisions based on future inputs.
    \item Our goal is to make the best possible decisions with the given information at each step.
\end{itemize}
Online algorithms are used in various real-world situations, such as routing internet traffic, managing computer memory, or making financial investments. They require efficient and adaptive decision-making because we cannot predict what is coming next.
A semi-online \cite{dwibedy2023semi} algorithm is like an intermediary  between online and offline algorithms. 

Imagine we are playing a game where we have to guess a secret number between 1 and 100. However, the game has a twist. We are allowed to know the first five numbers in advance, but after that, we have to guess the secret number as each new number is revealed.
In this example:

\begin{itemize}
    \item We have a bit of an advantage because we know the first five numbers.
    \item However, we still have to guess the secret number as each subsequent number is disclosed, without knowing the full sequence in advance.
    \item Our goal is to guess the secret number as accurately as possible using the initial information and adapting to the new numbers as they come in.
\end{itemize}

In this semi-online scenario, we have some prior knowledge (the first five numbers) to help us make better guesses, but we must adjust our strategy based on the new information as it unfolds.

We present three different frameworks such as offline, online and semi-online based on decision making as mentioned in figure 1

We used three colors in the figure viz green, pink and blue.  Green color indicates the input sequence to be processed, pink colour indicates the input sequence that is processed already, and  blue colour indicates the input sequence that we can see ahead.  So, in Offline cases  $\sigma_1$, $\sigma_2$ and $\sigma_3$ are the input sequences to be processed and all the input sequences are known from the beginning.  In online algorithm, first $\sigma_1$ is appearing on the fly which need to be processed.  Once, it is processed, now $\sigma_2$, appeared which need to be processed.  When $\sigma_3$ appeared, already $\sigma_1$ and $\sigma_2$ are processed.  In case of semi online algorithm, while processing $\sigma_1$, we know the next input sequence $\sigma_2$ and based on these two we have to process the input sequence.  So, from the above we can say that, in offline algorithm, we can see the whole input from the beginning, in online online algorithm only current input is visible where as in semi online algorithm along with current input we can see further input sequences depending on the situation.
\begin{center}
    \includegraphics[width=.4\textwidth]{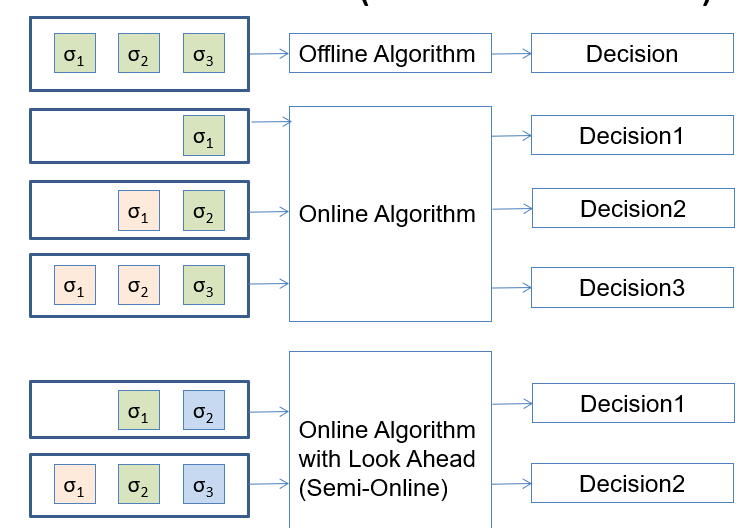}
\end{center}

\subsection{Competitive Analysis}
Competitive Analysis\cite{bachrach2002competitive,borodin2005online, boyar2007relative, breslauer1996competitive} is a method that involves comparing the costs incurred by an online algorithm when processing an input sequence to the costs of an optimum offline algorithm processing the same input sequence. An optimum offline algorithm is one that minimizes costs among all offline algorithms while handling the input sequence. Let's denote the cost of the online algorithm as \emph{ALG($\sigma$)} and the cost of the optimal offline algorithm as \emph{OPT($\sigma$)} for a given input sequence \emph{$\sigma$}. If the online algorithm serves \emph{$\sigma$} with a cost of \emph{ALG($\sigma$)}, it is considered c-competitive if there exists a positive constant b such that the following condition holds:
\begin{center}
$ALG(\sigma) \leq \emph{c} \cdot OPT(\sigma) + b$\\    
\end{center}

for every sequence of input \emph{$\sigma$}.  The benchmark for evaluating the effectiveness of online algorithm is the \emph{competitive ratio c} in this inequality.

Imagine we are responsible for calculating an examination scores for a group of students, and we have two different algorithms for calculating these scores: Algorithm P and Algorithm Q.  Algorithm P: This algorithm calculates the examination scores for students as their application forms and test scores arrive, one by one. It assigns scores based on the information available at the time of processing each student's application.  Algorithm Q: This algorithm has the advantage of knowing all the applicants and their test scores in advance. It can carefully review all the data and apply a more sophisticated scoring strategy to maximize the total score awarded.  Now, let's analyze the performance of Algorithm P in a competitive manner compared to Algorithm Q.

In the offline world (where we have complete knowledge), Algorithm Q may perform optimally, achieving the highest total examination score possible by carefully reviewing all the applicants and their scores.  In the online world, Algorithm P calculates examination scores as applications arrive, without knowing the quality of the remaining applicants. It makes decisions based on incomplete information. Competitive analysis compares the performance of Algorithm P against Algorithm Q in this online scenario.  The competitive ratio is used to measure how well Algorithm P performs compared to Algorithm Q in the online setting. It is defined as:

Competitive Ratio = (Total Examination Score Calculated by Algorithm P in the online scenario) / (Total Examination Score Calculated by Algorithm Q in the offline optimal scenario)

In our examination score calculation example, the total score calculated can be the sum of individual student scores, and the competitive ratio measures how Algorithm P's performance compares to the offline optimal Algorithm Q.  If Algorithm P has a competitive ratio close to 1, it means that it calculates examination scores nearly as well as Algorithm Q in the online setting. If the competitive ratio is much greater than 1, it indicates that Algorithm P's performance is significantly worse than the offline optimum.  Competitive analysis is relevant here because it helps us evaluate how well an online algorithm (Algorithm P) performs when it must calculate scores incrementally and without complete information about all the applicants and their test scores in advance.

\subsection{Practical Motivation}
Many applications, such as routing and transportation, scheduling, data structure organization, data transmission, packing, lot sizing, metrical task systems, and graph theory, have handled the online problem with LA.  Because there are more options for planning, adding LA to an algorithm should produce better outcomes.   It goes without saying that LA is useless without an algorithm that can use it. As a result, evaluating the value of LA and algorithm performance analysis are interdependent. Possible algorithmic faults cannot always be remedied later when it is discovered that a different choice would have been preferable due to the nature of sequential decision making under incomplete knowledge.  In the context of online optimization, lookahed is therefore considered a technique for enhancing an algorithm's power.

\subsection{Research Motivation}

Most often we wonder to know how computers make decisions on the fly, especially when there is no knowledge of future upcoming inputs to process  next?  Albers\cite{albers1998competitive} to raise the following challenging issue which is theoretically and practically significant.  If an online algorithm knows some
future requests as input in addition to the current request, what competitive
advantage can be gained?  In real life, we look ahead on future aspects in  situations like managing finances, handling logistics, or making decisions in healthcare. In design of efficient computer algorithms, without knowledge of future, we may not be  able to make best choices, thereby leading to poor performance and higher cost. The idea of using LA in algorithm and computer system design can be a game changer to improve performance and lower cost.

Our research is driven by the practical challenges faced in everyday decision making. We want to explore how giving algorithms a glimpse into the future can help them make better decisions. It's not just about theory; it's about finding ways for algorithms to plan ahead, adapt quickly, and potentially outperform others.  The advantages of LA have real-world applications. Algorithms becoming more efficient, using resources in a better way with flexibility in changing situations. By integrating LA, we hope to transform how decisions are made, not just on paper but in real applications.  With the advancement of technology we want to tap into different real life and computing applications to understand how LA can be practically applied across various areas.  Our research aims to uncover the benefits of LA in both real-life situations and computer applications. We want to show that giving algorithms the ability to see a bit into the future a practical step towards smarter decision-making in our fast-paced, ever-changing world.

\section{Look Ahead in Real Life}
We often make better decisions by visualizing the future. Based on the knowledge of future information, we can perform a task more efficiently by planning the steps of
action ahead to save time and cost.
The future may be one day, one month, one year or more than that.  Like the human decision making process, lookahead is a technique used in algorithms for economical usage of time and space.  It is applied by looking more inputs ahead before making a cost effective decision. Providing additional lookahead can enhance  performance by incorporating more information on inputs as a strategy for efficient algorithm design.  Planning ahead helps to boost the level of productivity and the efficiency of the outcome of a task. Through lookahead we gain additional information and knowledge which helps us to make better decisions to find efficient solutions.  We present some well known and interesting real life applications of lookahead as follows.
\subsection{Doctor's Clinic}

In real-life, the concept of Lookahead is seamlessly integrated into the functioning of a doctor's clinic during medical check-ups. Upon patients' arrival at the clinic, a systematic approach is adopted to streamline the process. At the reception counter, an organized system issues a serial number and a corresponding ticket to each patient. This simple practice serves as an efficient mechanism, enabling the medical staff and the attending physician to have a comprehensive Lookahead into the day's appointments.  By assigning a serial number and ticket, the clinic ensures that there is a structured order in which patients are examined. This measure facilitates a smooth workflow, allowing the medical professionals to anticipate and prepare for each patient's arrival. The doctor  works smoothly with the information provided by the serial number and ticket, gains valuable insight into the upcoming consultations, helping them plan and prioritize time effectively.  As a result, the Lookahead system implemented at the doctor's clinic not only benefits the healthcare providers but also enhances the overall patient experience. Patients can appreciate the organized approach, reducing waiting time and ensuring that they are addressed promptly. In this real-world application of Lookahead, the integration of such a system confirms how thoughtful planning can significantly optimize processes and improve the efficiency of medical services.

\subsection{Speed Cuber}
In the speed cubing, enthusiasts are well aware that achieving optimum solving time requires turning speed and lookahead. Turning speed refers to the rapid and precise execution of cube rotations during the solving process.  Lookahead in speed cubing involves the art of anticipating and strategizing moves by continuously scanning and identifying pieces while executing the current steps. This advanced skill demands a deep understanding of the cube's state. The synchronization of lookahead with turning speed is the major part of a speed cuber's success.  Lookahead, in particular, stands out as a crucial technique that sets the pace for speed cubers aiming to push the boundaries of their solving skills.  In essence, the relationship between lookahead and turning speed allows speed cubers to achieve optimal performance. 

\subsection{Stock Market Analysis}
In the dynamic world of stock market analysis, the lookahead emerges as a fundamental strategy employed by investors and traders to gain future stock prices.  Lookahead technique uses a comprehensive examination of historical price data and a study of market trends. The analysis of historical price data provides insights into the past performance of a stock, allowing investors to identify patterns, trends, and potential market behaviors. By using this information, investors can then project likely scenarios and outcomes involving the future stock prices.  Market trends play a crucial role in lookahead analysis, as they offer valuable clues about the trends of the market. This forward-looking perspective enables investors to position themselves strategically, aligning their portfolios with emerging trends.  In addition, technical indicators such as moving averages serve as powerful tools for lookahead analysis. By smoothing out price data over a specific time period, moving averages offer clear representation of trends and help to identify potential reversal points. Traders use this information to make informed decisions about when to buy or sell stocks, aligning their actions with the anticipated future movements of the market.  In essence, the application of lookahead techniques in stock market analysis provides a strategic  investment decisions. This decision applied by using lookahead always allows investors and traders to gain in the investments.

\subsection{Traffic flow predictions}
In transportation planning\cite{patzold2017look, powell2022designing}, the implementation of lookahead technique plays a vital role in predicting and managing traffic flow effectively. Forward-thinking allows transportation authorities  to anticipate potential traffic congestion  by using advanced technologies and data analytic. This enables them to actively plan and implement strategies for enhanced traffic management.  The traffic flow deals with the collection and analysis of real-time data from a network of cameras and sensors strategically placed throughout roadways. These sensors capture vital information about the volume, speed, and density of vehicle movement, creating a dynamic and comprehensive dataset. Through sophisticated algorithms and predictive modeling, transportation authorities from this data identify patterns, trends, and potential congestion points.  With the insights gained from lookahead analysis, transportation authorities can take timely and strategic actions to remove congestion and improve overall traffic flow. This may involve adjusting traffic signal timings for smoother vehicle movements. Additionally, in cases where congestion is unavoidable, authorities can reroute vehicles through intelligent traffic management systems, diverting traffic away from heavily congested areas to more efficient routes.  The real-time adjustments facilitated by lookahead techniques not only enhance the efficiency of traffic management but also contribute to a reduction in travel times, fuel consumption, and environmental impact.  In essence, the utilization of lookahead in traffic flow predictions represents a shift in transportation planning.

\subsection{Inventory management}
In business operations, the strategic utilization of lookahead techniques proves optimizing inventory management. Enterprises uses the power of foresight by analyzing past data to forecast future demand accurately.  The foundation of this forward-looking approach is based on a thorough examination of historical data by using factors such as sales patterns, seasonal fluctuations, and market trends. Through a proper analysis, businesses gain valuable insights into the consumer demand and employ predictive modeling to anticipate future requirements.  The 
 first objective of lookahead in inventory management is to prevent overstocking of valuable capital and warehouse space. The second objective is to avoid stock-outs that can result in lost sales and dissatisfied customers. By accurately predicting future demand, businesses can make a required balance, ensuring they have the right amount of stock on hand to meet customer needs while minimizing excess inventory costs.  In essence, the application of lookahead in inventory management applies with supply chain optimization. Businesses that use this methodology not only enhance their operational efficiency but also cultivate a competitive edge by consistently meeting customer demands, reducing carrying costs, and creating a responsive business model.

\subsection{E-commerce recommendations system}

In e-commerce, the integration of lookahead techniques is a stepping stone in the development and effectiveness of recommendation systems. Online retailers use the power of anticipation by analyzing users' browsing and purchase history.  They use this data to predict and recommend products as per individual preferences.  The recommendation system begins its work by comprehensively examining the user's interaction with the platform. This includes scrutinizing the user's browsing history. viewed products and past purchase behavior. By extracting patterns and trends from this historical data, the system can find the user's preferences, interests, and future needs.  Furthermore, the incorporation of machine learning algorithms enhances the use of the lookahead process. These algorithms continuously learn and adapt, refining their predictions over time as users engage with the platform. The result is a recommendation system that becomes increasingly accurate and personalized, providing users with suggestions that is more closely associated with their evolving tastes and preferences. 
 In essence, the use of lookahead in e-commerce recommendation systems represent a strategic effort to enhance the user experience and drive engagement. By accurately anticipating user preferences and suggesting relevant products, online retailers not only increase the likelihood of successful conversions but also cultivate a more personalized and customer-centric shopping experience, with increasing loyalty and satisfaction among their user base.
\subsection{Organizer and calendar}
In the case of personal and professional scheduling, the role of lookahead is applied through the integration of calendar applications such as Google Calendar and Microsoft Outlook, as well as various task management tools and organizers. Lookahead, in this context, enables individuals to plan and prioritize tasks and deadlines with foresight and efficiency.  Calendar applications such as Google Calendar and Microsoft Outlook, harness the power of lookahead to offer users a comprehensive view of their upcoming schedule by allowing users to input events, appointments, and deadlines. These applications facilitate a proactive approach to time management. Task management tools and organizers further use lookahead techniques to enhance productivity. By analyzing task lists, deadlines, and project timelines, these tools empower users to prioritize their activities based on a forward-looking perspective. The ability to anticipate upcoming tasks and deadlines allows individuals to allocate time and resources effectively thereby reducing the likelihood of last-minute rushes or missed deadlines.  In essence, the incorporation of lookahead in calendar applications, task management tools, and organizers signify how individuals approach time management. By providing a proactive and organized framework, these tools empower users to not only meet their immediate responsibilities but also plan ahead, by allowing a sense of control and efficiency in both personal and professional endeavors.

\subsection{Disease prediction}
In the public health, the application of lookahead concept represents a vital approach to disease prediction. Public health agencies find out the patterns underlying the emergence and spread of diseases with the help of advanced technological tools and analytical methodologies to. This analysis includes an examination of symptoms, geographical locations, and historical disease patterns upon which predictive models are constructed.  In public health, experts use a smart method called lookahead to diagnose diseases. They look for relevant information, like symptoms, demography of population and historical information on diseases. Based on relevant advance information, it can be determined how diseases might spread for the further treatment and recovery. Looking ahead helps them in advance preparation to reduce or stop spreading of the diseases for keeping the community safe.  With better technology and advance tools the prediction and prevention of diseases can be ensured at the earliest stage.

\subsection{Natural Disaster Management}
To handle and manage natural disasters such as floods, wildfires, volcanoes, and extreme weather conditions, special agencies use LA models. The models are used to predict the expected time and location of occurrence of such disasters. For example, weather radar and rainfall data are used to predict the likelihood of heavy rain for causing of unexpected floods. Emergency warnings and weather alerts are issued to ensure public safety. 
Proactive measures can be taken ahead of time by monitoring the potential threats of wildfires caused by strong winds, high temperatures, and low humidity. To prevent rapid spreading of fire, firefighters and equipment can be placed in appropriate locations apriori.  

\subsection{Space mission}
Space agencies such as NASA and ISRO utilize LA approaches to carefully plan and execute space missions. LA is an essential concept in this planning process, involving the consideration of future events and developments from the inception of the mission.  When planning a space mission, agencies use LA models to anticipate expected possibilities of consequential events in the future.  The events include predicting the positions of planets, the availability of certain resources, and the best time to launch spacecraft to reach respective destination.  By using LA, space agencies can optimize the launch time of  spacecraft through the most efficient and optimal route.  LA in space mission involves the long-term objectives like exploring a distant planet, studying a specific concept, or conducting experiments in space. In essence, the use of LA in space mission reflects a commitment of thorough preparation and foresight. As technology advances, the LA concept continues to evolve, enabling space agencies to push the limits of exploration and expand the understanding of the universe.

\subsection{Video games}
In video games, LA is a smart technique used by computer-controlled enemies. These enemies can guess beforehand what players might do next, making the game more fun and challenging.  The next moves are predicted and responded dynamically by the enemy players against a virtual bad guy.  Game developers use LA to make video games feel more real and exciting by making it dynamic with unpredictable experience. As technology gets better, these smart features in video games keep evolving, making the enemies in games smarter and the game play more engaging for players.

\subsection{Self-driving cars}

Self-driving cars, also known as cars that can drive themselves, use LA to stay safe on the roads. Special sensors are used  to  predict the presence of objects and occurrence of events in the surrounding.  In real life, LA is like a superpower for self-driving cars which helps them plan where to go, make good decisions, and use their resources wisely. For example, the car's sensors check how fast other cars are going and figure out if there are any obstacles by looking at the whole traffic situation. This helps the self-driving car to decide the best way to drive, following the rules and keeping everyone safe.  LA makes the self-driving car really smart at making quick decisions when something unexpected happens, like a car suddenly stops in front, then the car can slow down or change lanes to avoid problems.  In simple terms, LA is like having a smart helper in the self-driving car that looks into the future, plans the best route, and ensures safety.

\section{Look Ahead in Computing}
LA is used in various areas of computing to optimize algorithms and improve performance. Search and decision-making algorithms use LA as a mechanism for increasing the power of algorithms.  We present the role of LA in ten well known and widely used optimisation computational problems as follows.  

\subsection{List Update Problem}
In recent years, there has been a growing fascination towards the study and exploration of the competitive analysis\cite{bachrach2002competitive, borodin2005online, boyar2007relative} of online algorithms. Researchers have delved into various online problems across domains such as resource allocation, data structures, graph problems, scheduling, and navigation. The list update problem holds theoretical and practical significance in the area of online algorithms and self-organizing data structures\cite{hester1985self}.

The list update problem\cite{rakesh20015thesis, simha1994lookahead,albers1998competitive, sleator1985self} revolves around efficient access of a set of items arranged in an unsorted linear list. Starting with an initial list of $n$ items, a list update algorithm must access and reorganize a sequence of requests of size $m$, where $m \geq n$. Each request specify an item in the list. To access a request in the list, the algorithm linearly searches through the list, starting from the front, until it finds the desired item. The cost of accessing the $i^{\text{th}}$ item in the list is $i$ according to standard cost model.

Following a request, the accessed item can be moved to a closer position to the front without incurring extra cost, known as a free exchange. An exchange between two consecutive items in the list, excluding free exchanges, incurs unit cost  and is termed as paid exchange. 
The primary objective is to process the request sequence in a way that minimizes the total access and reorganization cost. An online list update algorithm achieves this goal by serving each request without the knowledge of future requests.

An online algorithm for the list update problem performs better with the knowledge of the current request and future upcoming requests. In real-world, requests come in blocks of different sizes, and they can be generated faster than the algorithm can handle them, causing a backlog. Sometimes, it's possible to delay serving of requests to wait for others. It is interesting and challenging to explore that "what it is worth to know the part of the future?"  

Competitive Analysis with LA provides a measure of how well the online algorithm performs in comparison to the optimum offline algorithm while serving various request sequences.

\subsection{Ski Rental Problem}

The Ski Rental problem is an illustration for online rent or buy problem. In the ski rental problem, an individual visits to a ski resort without knowledge of the usage duration of skiing adventure.  The ski can be either rented with cost $r$ or can be bought with cost $b$.  If the ski is used for a short duration more frequently, then it is wise to  rent. Otherwise, it is wise to buy for long term usage of the ski. If we know the specific period of future usage of the ski before hand using LA then we can make smarter decision on rent or buy to minimize the cost.

\subsection{Routing Problem}

A specialized network, known as a Delay Tolerant Network (DTN)\cite{allulli2005power, allulli2008power, xian2011look}, is designed to operate effectively in challenging environments and over extensive distances, such as in space communications. Conventional internet systems struggle to function under these conditions, often experiencing frequent outages, high error rates, and prolonged delays lasting hours or even days. DTNs offer a solution to these challenges, ensuring reliable information transmission and interconnectivity.  First introduced in 2003, DTN address the unique difficulties encountered in space communication, where the vast distances between Earth and orbiting spacecrafts lead to inevitable delays, interruptions, errors, and data losses. These issues need to be handled at the application level due to the fact that as existing terrestrial networking technologies are inadequate in resolving such challenges.  

Routing poses a frequent challenge in DTNs due to the absence of a continuous connection. Existing routing systems for DTNs provide best-effort services but struggle to enhance Quality of Service (QoS) or prioritize messages. In the ALARMS (Look-Ahead Routing and Message Scheduling) method, Powell \cite{powell2022designing},  improved QoS in DTNs by utilizing more precise information about various routing parameters. Here it is assumed that the presence of ferry nodes are present following predetermined routes to communicate with the gateway node at each location along the path. Additionally, the system transmits look-ahead routing data to the gateway nodes, detailing the expected arrival and stay times at each gateway node along the path for the next two cycles. Gateway nodes use this information to predict the delivery delay of each message when sent via different ferries.

\subsection{Bin Packing Problem}

Bin packing problem\cite{ahlroth2010power,grove1995online} is another computational problem where LA is applied. In this problem, a set of items, each with a specific size (or weight), must be packed into a finite number of containers (or bins) of a fixed capacity\cite{ahlroth2010power},\cite{grove1995online}. The objective is to minimize the number of containers used or, equivalently, to maximize the space utilization within the containers. Imagine a loading dock where items arrive, and our task is to load each item into trucks on arrival. Each truck has a maximum weight capacity, and in this scenario, an item's weight is equivalent to its "size." We follow a one-dimensional bin packing approach, where once loaded into a truck, items are never taken out.

In typical online bin packing, there is an infinite number of trucks accessible throughout the algorithm, allowing for flexibility. On the other hand, in bounded-space bin packing, only a predetermined number of vehicles can fit. Some trucks may need to be closed and sent away to make room for new, empty vehicles. Once closed, a truck cannot have any additional items loaded onto it. 
An alternative strategy is to introduce a warehouse with a fixed capacity. Instead of loading items onto trucks immediately, if there is space, items can be stored in the warehouse. This approach simplifies the development of an optimum algorithm that always has one truck available. However, for a more optimal algorithm with a finite look ahead, the process becomes more complex, involving multiple trucks.

The goal is to optimize the overall cost in the future rather than focusing on a specific number of future items. To maximize the use of the look ahead, it's crucial to carefully assess the cost of processing an item during bin packing. The challenge with the presented bin packing technique is that an item may remain unused for an extended period, requiring an infinite look ahead as it depends on subsequent items in the input sequence. 
While this may be acceptable, considering that even typical online bin packing algorithms may leave items in open bins indefinitely, it's interesting to note that the straightforward algorithm can be transformed into an online algorithm using a bounded look ahead.

\subsection{Scheduling Problem}

In the job scheduling problem\cite{coleman2004lookahead, dwibedy2023semi} with two machines, denoted as $\emph{M}_{1}$ and $\emph{M}_{2}$, tasked with scheduling $\emph{n}$ jobs, each job has an associated arrival time $\emph{r}_{j}$ and a processing time $\emph{p}_{j}$. The objective is to minimize the total completion time, denoted as $\emph{c}_{j}$, which represents the sum of completion times for all $\emph{n}$ jobs. 
The algorithm operates with continuous access to the arrival times and processing specifications of the upcoming job, as well as the backlog of tasks awaiting completion. Unlike traditional scheduling methods, this algorithm doesn't delay decisions between scheduling and waiting. It evaluates each job in the wait queue promptly.

At any given time \emph{t}, if machine $\emph{M}_{1}$ is idle and machine $\emph{M}_{2}$ is expected to become idle at time $\emph{t}^{`}$, the algorithm assesses the current set of jobs, denoted as $\emph{J}_{1}$  through $\emph{J}_{i}$ in the wait queue. Simultaneously, it considers a job, $\emph{J}_{i+1}$, in the lookahead queue, set to arrive at time $\emph{r}_{j+1} > t$. The algorithm's task is to decide whether to execute job $\emph{J}_{1}$ immediately or to wait for job $\emph{J}_{i+1}$ to arrive.
A crucial observation in the algorithm is that even-indexed jobs are scheduled in order on one machine, while odd-indexed jobs (excluding $\emph{J}_{1}$) are scheduled in order on the other machine. This pattern, stemming from scheduling jobs in ascending order of processing time, enables the algorithm to make decisions without the need to calculate the entire schedule, streamlining the decision-making process.

\subsection{Caching Problem}
Look-ahead caching\cite{hellstrom2008design}, commonly known as prefetching, is a strategy employed in caching systems to enhance performance. Instead of waiting for users or applications to request items from the cache, look-ahead caching predicts potential future needs and proactively loads these items into the cache before they are actually requested. This proactive approach is designed to minimize delays and optimize the overall efficiency of the cache.

The caching system continuously monitors the way users or applications interact with the cache, taking note of usage patterns, access frequencies, and any observable sequential trends in their requests.  Based on the observed patterns, the system utilizes various algorithms to predict which items are likely to be accessed in the near future. These predictions consider factors such as historical access data, temporal and spatial locality, and other statistical methods.  
Once the system identifies items anticipated to be needed soon, it fetches and loads them into the cache ahead of time, even before actual requests are made. This ensures that when users or applications do request these items, they are already available in the cache, reducing response times and minimizing instances of cache misses.  
The system adapts over time by incorporating new access patterns into its predictive models, adjusting its strategies for prefetching. This adaptive process may involve assigning more significance to recent accesses or considering more intricate patterns.

While look-ahead caching offers significant performance improvements, there are trade-offs to consider. Incorrect predictions may lead to inefficient use of cache space and, in some cases, degrade overall performance. Additionally, the predictive and prefetching processes introduce some overhead, potentially impacting the responsiveness of the entire system.  Various strategies and algorithms are employed in look-ahead caching. Sequential Prefetching focuses on preparing items that are expected to be accessed in a sequence. Temporal Locality emphasizes prefetching items that have been frequently accessed in the recent past. Spatial Locality prefetches items that are spatially close to the currently accessed item. Machine Learning-based Approaches leverage advanced techniques to predict future accesses, often training models on historical access patterns.  User-Based Approaches personalize prefetching based on individual user behavior, utilizing user profiles or historical data if available.  Look-ahead caching proactively manages cache content by predicting future user or application behavior. The goal is to provide a more responsive experience by minimizing the time associated with waiting for items to be fetched from the cache.

\subsection{Paging Problem}
Look-ahead in paging refers to a technique used in computer memory management, specifically in virtual memory systems that utilize paging, to anticipate and prefetch pages into main memory before they are actually needed. This strategy aims to reduce the occurrence of costly page faults and enhance the overall performance of the system\cite{breslauer1996competitive}. 

The operating system monitors the patterns of page accesses made by running programs. It keeps track of which pages are being accessed and the sequence in which these accesses occur.  Based on the observed access patterns, the operating system employs various algorithms and heuristics to predict which pages are likely to be accessed in the near future. These predictions are often guided by principles of temporal and spatial locality, which suggest that recently accessed pages are likely to be accessed again soon and that neighboring pages tend to be accessed together.  Once the likely pages to be accessed in the near future are predicted, the operating system proactively loads these pages from secondary storage (like a hard disk) into the main memory before they are actually requested by the executing program. This step is called prefetching.  By prefetching pages in advance, the system aims to avoid or reduce page faults. Page faults occur when a program accesses a page that is not present in the main memory, causing the operating system to fetch the missing page from secondary storage. This operation is time-consuming and can significantly slow down program execution. Prefetching helps mitigate this delay by making sure that frequently accessed pages are already in memory, thus reducing the likelihood of page faults.  Look-ahead paging strategies may evolve over time. The operating system can adjust its predictions and prefetching techniques based on changing access patterns. Strategies can range from simple approaches like sequential or adjacent page prefetching to more complex algorithms that use historical data and statistical models to predict future accesses.  While look-ahead paging can be effective in reducing page faults and improving performance, there are trade-offs to consider. Prefetching too aggressively might lead to inefficient use of memory if the prefetched pages are not actually used. Additionally, there's an overhead associated with prediction and preloading, which might impact the overall responsiveness of the system.

LA in paging is a proactive approach to memory management where the operating system predicts which pages are likely to be accessed in the future and fetches them into main memory before they are explicitly requested. This strategy leverages access pattern analysis to optimize memory usage, minimize page faults, and enhance the overall performance of programs running in a virtual memory environment.\\

\subsection{Graph Coloring Problem}
In an online algorithm, A, the task is to promptly assign a workable color to vertex $\emph{v}_{t}$, ensuring that this color differs from those already assigned to the previously revealed neighbors of $\emph{v}_{t}$. The primary objective is to minimize the overall count of colors utilized\cite{albers2021tight}.

The Graph Coloring Problem (GCP) is a well-explored combinatorial challenge in graph theory. It involves assigning colors to the vertices of a given graph in such a way that no two connected vertices have the same color. Online graph coloring, on the other hand, presents a classic problem that spans both graph theory and online computation. In this context, the goal is to efficiently assign colors to vertices as they are presented, taking into account the connectivity between vertices. The challenge lies in making color decisions on-the-fly as new vertices are introduced, aiming to adhere to the coloring constraints while adapting to the evolving structure of the graph.  

A problem instance is characterized by an undirected graph G = (V, E), comprising a set of vertices V and a set of edges E. Let the cardinality of V be denoted as $|V|$ = n. The vertices arrive sequentially in a sequence $\sigma$ = $\emph{v}_{1}$, . . . ,$\emph{v}_{n}$, and the order of this sequence may be influenced by an adversary. As each new vertex $\emph{v}_{t}$ arrives, where $1 \leq t \leq n$, the edges connecting it to previously arrived vertices $\emph{v}_{s}$ with $s < t$ are unveiled.  As each new vertex $\emph{v}_{t}$ arrives, where $1 \leq t \leq n$, the edges connecting it to previously arrived vertices $\emph{v}{s}$ with $s < t$ are revealed. This sequential unveiling of edges provides a dynamic aspect to the problem, where the connectivity of the graph evolves with the introduction of each new vertex in the specified sequence.  In the context of a graph G, A(G) represents the count of colors utilized by algorithm A, while $\chi(G)$ stands for the chromatic number of G, denoting the minimum number of colors required to offline-color G. An online algorithm A is considered c-competitive if, for any given graph G, the inequality $A(G) \leq c \cdot \chi(G)$ holds true.

An online algorithm, A, is said to have a lookahead of length l if, when vertex $\emph{v}_{t}$ arrives, the algorithm can also access information about the subsequent l vertices $\emph{v}_{t+1}$, . . . , $\emph{v}_{t+l}$, including their connections to vertices within the set {$\emph{v}_{1}$, . . . , $\emph{v}_{t+l}$}. Alternatively, an algorithm may employ a buffer with a capacity of $b$, allowing it to temporarily store vertices. The primary requirement is that by the end of step $t$, the algorithm must have colored at least $t−b$ vertices. A buffer offers more versatility compared to lookahead as it enables the algorithm to partially reorder the input sequence and defer coloring decisions.

\subsection{Graph Matching Problem}

In the field of graph theory and computer science, addressing the graph matching problem involves the task of identifying connections or mappings between nodes in multiple graphs. This challenge has practical applications ranging from image analysis to biological data interpretation and social network analysis. Exploring the use of look-ahead techniques in this context reveals potential improvements in the accuracy and efficiency of correspondence identification.

To illustrate, imagine comparing two images in image analysis to identify similarities in object placement. The graph matching problem in this scenario entails finding correspondences between key points or features in the images. By applying look-ahead techniques, the algorithm can predict potential connections between these key points before they are explicitly revealed. This anticipatory approach may involve forecasting the likely sequence of identifying key points, estimating potential edges between them, and dynamically adjusting the matching strategy as more features are unveiled.

In practical terms, incorporating look-ahead not only optimizes existing matching algorithms but also considers the evolving structure of the images. It involves estimating feature similarity in advance, guiding the algorithm to prioritize potential matches and refine the matching process as the analysis progresses. The feedback mechanism, highlighted as beneficial, emphasizes continuous learning, where the algorithm refines predictions based on actual correspondences identified in the images over time.  Applying look-ahead techniques proves advantageous in addressing the graph matching problem, offering a more informed and efficient approach, particularly in scenarios such as image analysis.

\subsection{Secretary Problem}
Consider a manager who wishes to select the $n^{\text{th}}$ applicant who will make the best secretary.  One by one, in sequence, the candidates are interviewed. After the interview, a decision of selection or rejection must be made regarding each individual applicant.  An applicant cannot be recalled again after being rejected.  During the meeting, the manager acquires adequate data and complete information to rank the candidates instantly without knowledge of the performance of future candidates to be interviewed.  This problem is popularly called as Secretary Problem.  Besides this the problem is also known as “the optimal stopping problem”, “the marriage problem”, “the best choice problem”.  Here, it is too difficult to apply look ahead to find out the best out of alternatives.

\section{Taxonomy of Lookahead}
There are three bases on which we can define Lookahead.  We will discuss types of Look Ahead on the basis of pattern of input, size of the input and presence of input.  Lookahead in list update problem can be studied by using two different models of lookahead i.e. weak lookahead and strong lookahead. 

Let $\sigma$ = $\sigma(1)$, $\sigma(2)$,......,$\sigma(m)$ be a request sequence of length m.  $\sigma(t)$ denotes the request at time \emph{t}.
For a given set S, card(s) denotes the cardinality of s, let \emph{l} $\geq$ \emph{1} be an integer.
\subsection{Weak Lookahead}

Weak lookahead of size \emph{l} : The online algorithm sees the present request and the next \emph{l} future request.  More specifically, when answering $\sigma(t)$, the online already knows $\sigma(t+1)$, $\sigma(t+2)$, ...., $\sigma(t+l)$.  However, requests $\sigma(s)$, with \emph{s} $\geq$ \emph{t+l+1} are not seen by the algorithm at time t.\cite{albers1998competitive} 
\subsection{Strong Lookahead}
Strong lookahead of size \emph{l} : The online algorithm sees the present request and a sequence of future requests.  This sequence contains \emph{l} pairwise distinct items, which also differ from the items requested by the present requests. More precisely, when serving request $\sigma(t)$, the algorithm knows requests $\sigma(t+1)$, $\sigma(t+2)$,...., $\sigma(t')$, where \emph{t'} = min {\{\emph{s} $>$ \emph{t} $|$ card({$\sigma(t)$, $\sigma(t+1)$,....,$\sigma(s)$ = \emph{l+1}}}\} \cite{albers1998competitive}.
\subsection{Static Lookahead}
Albers \cite{albers1998competitive}  considered fixed size static look ahead and static optimum
offline algorithms for analyzing the performance of online list update algorithms. In
static lookahead, the online algorithm foresees a constant size lookahead l of the future
requests at each step.
\subsection{Dynamic Lookahead}
In the dynamic look ahead, the online algorithm can foresee a lookahead of $\emph{l}_\emph{v}$ of the future requests whose size is not fixed in each step. Here, $\emph{l}_\emph{v}$ = \emph{i} -1, \emph{i} is the position of the list for the currently accessed item of the request sequence. \cite{rakesh20015thesis} 

\subsection{Short Lookahead}

Short lookahead, in both real-life applications and computing, involves a limited or brief anticipation of events or outcomes within a system. This concept revolves around considering only a small, immediate window into the future, typically just a few steps ahead, rather than attempting to predict events far into the future. The essence of short lookahead lies in its application across various fields where quick decisions or process optimization are paramount, without the need for deep, long-term predictions.

In the context of gaming, short lookahead is crucial in strategies that involve evaluating just a few moves ahead to determine the optimal move in a game. For instance, chess engines may consider only a handful of moves due to the complex branching factor of possible moves. Similarly, in scheduling tasks or processes, short lookahead prepares planning and decision-making based on immediate upcoming events or tasks, without considering a long-term schedule. This approach is frequently employed in real-time scheduling systems to enhance efficiency.

In control systems and control theory, short lookahead predicts the immediate future behavior of a system to make adjustments to a control signal. This practice helps in improving stability and performance in feedback control systems. Moreover, when optimizing a process or algorithm, short lookahead involves making decisions based on a few upcoming steps to adapt changing conditions, rather than planning far ahead.  

In navigation algorithms, short lookahead plays a pivotal role by considering only the immediate path ahead. This is especially useful for avoiding obstacles or making navigation decisions in real time. The philosophy of short lookahead is to create a balance between making prompt decisions that respond to immediate conditions and making accurate predictions that demand more computation and time. Short lookahead, therefore, emerges as a versatile approach that finds applicability across diverse domains, facilitating agility and efficiency.

\subsection{Medium Lookahead}

Medium lookahead, a versatile strategy applicable in both real-life scenarios and computing, involves the consideration of events, trends, or outcomes within a timeframe that falls between short-term and long-term predictions. This forecasting technique strikes a balance between immediate decisions and those with more distant involvement, allowing decision-makers to navigate a middle ground that encompasses both short-term demand and longer-term planning.

In the domain of project management, a medium lookahead approach proves utmost important. It entails planning and anticipating events and milestones that extend several months to a few years into the future. By doing so, it aids in resource management, deadline setting, and aligning project goals with broader organizational objectives. Similarly, in marketing and business strategy, a medium lookahead strategy unfolds over a period of one to three years, involving the forecasting of market trends and consumer behaviors. This foresight enables businesses to adapt their marketing strategies in changing market conditions effectively.

Energy planning also benefits from a medium lookahead perspective, enclosing the prediction of energy consumption patterns, availability of renewable resources, and energy demand trends over several years. This informs critical decisions about energy infrastructure and resource allocation. Likewise, in urban planning, a medium lookahead strategy involves anticipating population growth, infrastructure needs, and land use changes over a span of five to ten years. This foresight informs decisions regarding zoning, transportation, and overall city development.

Healthcare planning adopts a medium lookahead approach by forecasting healthcare needs, disease trends, and infrastructure requirements over a timeframe of a few years. This strategic foresight aids in resource allocation and the thoughtful planning of medical facilities. The medium lookahead approach, therefore, emerges as a strategy that balances short-term tactical decisions with long-term strategic planning. It equips decision-makers to consider both immediate needs and the potential consequences of their actions over a more extended period, proving particularly effective in situations that demand acceptance to changing conditions while considering the medium-range timeline.
\subsection{Long Lookahead}

Long lookahead, a strategic approach applied both in real-life scenarios and computing, involves an extensive anticipation or prediction of events, outcomes, or consequences within a system. Unlike short or medium lookahead, long lookahead dive into the future events, providing decision-makers with a broader perspective on future possibilities and trends. This concept finds application across various domains where making informed decisions requires considering a more extended view of future developments.

Financial planning benefits significantly from a long lookahead perspective, especially in financial analysis. Here, the approach involves projecting financial trends and outcomes over an extended period, often spanning several years likely ten to fifteen years. This assists individuals and businesses in making decisions aligned with long-term financial goals. Similarly, in climate science, long lookahead plays a crucial role in modeling and simulating climate patterns and changes over decades or even centuries. This enables a deeper understanding of the long-term impact of factors such as greenhouse energy emissions.

Supply chain management also embraces a long lookahead approach, where forecasting demand and supply trends over a substantial timeframe optimizes production, distribution, and inventory management. In the context of urban planning, decisions related to infrastructure and city planning are made with a long lookahead, considering their effects over many years to ensure sustainable growth and development. Additionally, in research and strategic planning, a long lookahead involves projecting technological, societal, or industry trends far into the future to guide research and innovation efforts.

The ability to anticipate challenges and opportunities over an extended period enhances the comprehensiveness of decision-making, making long lookahead a valuable strategy in contexts where a more profound understanding of future developments is essential.

\section{Research Challenges and Future Directions}

Intuitively, we observe that LA can be used as a powerful tool and framework for future researchers in design of efficient computational models and algorithms for solving non-trivial and challenging optimization problems. Using the concept of LA, We identify and present the following interesting and non-trivial research challenges as future potential research directions.

\begin{itemize}
    \item Development of novel temporal lookahead models to solve real life and computational problems
    \item Design of specific classes of lookahead models based on real life applications
    \item Design of bounded size lookahead models with upper bound and lower bound on the size of the input for improving the performance of algorithms and computing systems
    \item Design and analysis of new paradigms and algorithms based on lookahead models for efficiently solving real world problems
     
\end{itemize}



\end{document}